\documentclass[sigconf]{acmart}
\AtBeginDocument{%
  \providecommand\BibTeX{{%
    \normalfont B\kern-0.5em{\scshape i\kern-0.25em b}\kern-0.8em\TeX}}}

\copyrightyear{2023}
\acmYear{2023}
\setcopyright{rightsretained}
\acmConference[CHI EA '23]{Extended Abstracts of the 2023 CHI Conference on Human Factors in Computing Systems}{April 23--28, 2023}{Hamburg, Germany}
\acmBooktitle{Extended Abstracts of the 2023 CHI Conference on Human Factors in Computing Systems (CHI EA '23), April 23--28, 2023, Hamburg, Germany}\acmDOI{10.1145/3544549.3573855}
\acmISBN{978-1-4503-9422-2/23/04}
\begin{document}

\title{Location-based AR for Social Justice:\\
Case Studies, Lessons, and Open Challenges}

\author{Hope Schroeder}

\email{hopes@mit.edu}
\affiliation{%
  \institution{MIT}
  \country{USA} 
}

\author{Rob Tokanel}
\email{tokanel.r@gmail.com}
\affiliation{%
  \institution{Columbia University}
  \country{USA} 
}

\author{Kyle Qian}
\email{kylecqian@gmail.com}
\affiliation{%
  \institution{Stanford University}
  \country{USA} 
}

\author{Khoi Le}
\email{khoivietle2@gmail.com}
\affiliation{%
  \institution{Stanford University}
  \country{USA} 
}

\renewcommand{\shortauthors}{Schroeder, Tokanel, Qian, et al.}

\begin{abstract}
  \textit{Dear Visitor} and \textit{Charleston Reconstructed} were location-based augmented reality (AR) experiences created between 2018 and 2020 dealing with two controversial monument sites in the US. The projects were motivated by the ability of AR to 1) link layers of context to physical sites in ways that are otherwise difficult or impossible and 2) to visualize changes to physical spaces, potentially inspiring changes to the spaces themselves. We discuss the projects' motivations, designs, and deployments. 
  We reflect on how physical changes to the projects' respective sites radically altered their outcomes, and we describe lessons for future work in location-based AR, particularly for projects in contested spaces.

\end{abstract}




\maketitle

\section{Introduction}

Augmented reality (AR) adds layers of content onto our perceptual field that can meaningfully alter the way we see and interpret the world around us. It can help us examine physical spaces by adding location-specific context and imagine how the world could be different. AR allows us to experience the impossible, forging new ways of interacting with our surroundings through an exciting emerging medium. These capabilities become especially powerful when applied to contested public spaces, which are often the sites of competing worldviews and visions for the future.

Physical symbols of the legacy of slavery in the US\textemdash such as Confederate and antebellum period monuments\textemdash have become the subject of increased scrutiny and debate over the past several years. In the wake of the mass shooting in Charleston in 2015, and in the years following the white supremacist rally in Charlottesville in 2017, calls to investigate the legacy of monuments intensified, and extended beyond Confederate monuments to other contested public spaces as well. 

We were a team of students based at Stanford and Columbia interested in how carefully designed AR experiences could help interrogate structures and power dynamics in public space. We believed we could add layers of storytelling that brought relevant voices to the space in ways that would be impossible without the assistance of technology. We received a Magic Grant from the \href{https://brown.columbia.edu/}{Brown Institute for Media Innovation} for \textit{Charleston Reconstructed}, a project intervening on Charleston, South Carolina's most infamous monument using AR. A local prototype for \textit{Charleston Reconstructed} became its own project at Stanford called \textit{Dear Visitor}.

In our first project, \textit{Dear Visitor}, a monument's conspicuous \textit{absence} was an opportunity to not only instantiate the monument as it should have been, but also to digitally add layers of institutional memory that are impossible through purely traditional media. In a fitting twist, the AR project eventually led to a permanent physical change to the space.

In our second project, \textit{Charleston Reconstructed}, a monument's conspicuous \textit{presence} highlighted the legacy of slavery in Charleston, South Carolina's most prominent park, Marion Square, for decades. We used AR to enact three interventions and surface historical perspectives in the square. We completed and beta-tested the project, but the project lagged physical changes in the world when the original monument was taken down in June 2020 in the wake of the George Floyd protests.

In this piece, we describe the designs of the projects, lessons learned from their deployment, and open challenges we still see for future work of this kind.

\subsection{Technological motivation}

Ronald Azuma has asserted that the fundamental value of AR comes from the unique combination of digital content and physical reality such that ``virtual content is connected to reality in compelling and meaningful ways." \cite{azuma} AR has seen myriad technological advances in recent decades and several surges in popularity in the commercial sphere. However, AR has yet to live up to its potential as a vehicle for facilitating meaningful experiences situated in the physical world. Though modern AR systems are adept at detecting basic properties such as position and shape, they generally lack the ability to infer physical properties like material, or semantic properties like identity and cultural significance. In other words, modern AR systems can understand \textit{where} things are, but seldom \textit{what} they are, and rarely \textit{why} they matter.

As a result, Pokémon Go, the most well-known AR application in popular use, features content which is overlaid on reality, but which lacks meaningful connection to it. Other platforms, such as Google Lens, are capable of basic semantic understanding of an object or site, but are currently primarily reference tools that lack interactivity and content.

While the technology required for generalized semantic understanding using commercially available AR does not yet exist, many necessary elements for creating location-based AR experiences do. We used visual markers to trigger the location-based experiences we designed, thus deploying a temporary solution to the problem of semantic understanding of the space. This allowed us to explore the promises and challenges of creating meaningful location-based AR before it is achievable at scale.

With both technological and social goals in mind, we created \textit{Charleston Reconstructed} and \textit{\href{https://www.dearvisitor.app/}{Dear Visitor}}, two location-based AR experiences. Lessons from these case studies will be useful in planning for future projects as the technology develops. 

\subsection{Related work}

Several other projects have meaningfully engaged with site-specific AR, such as \textit{110 Stories}, which used AR to visualize an outline of the Twin Towers against the New York skyline, showing the towers' outline before the 9/11 attacks. The experience demonstrates an intentional usage of AR which combines content with context, elevating both the physical space and the digital augmentation. Another project, \textit{Stonewall Forever}, incorporated archival materials related to the Stonewall riots into an AR smartphone app which projected a large, dynamic rainbow into the physical site of the riots in New York City, and invited users to leave comments on the virtual monument. Another project currently in development called ``If These Streets Could Talk" will show elements of Jewish archival history overlaid on sites in the Jewish quarter of Budapest, with planned deployments in other European cities as well.

\section{Case Studies}

\subsection{Dear Visitor}

\subsubsection{Motivation for the project}

In 2015, Chanel Miller was sexually assaulted behind a dumpster on Stanford campus. Her \href{https://www.buzzfeednews.com/article/katiejmbaker/heres-the-powerful-letter-the-stanford-victim-read-to-her-ra}{victim impact statement} was published on BuzzFeed and went viral during the \#MeToo movement. As part of its handling of the sexual assault case, Stanford agreed to build a contemplative garden at the site of the assault per Chanel Miller's request, which would include a plaque with a quote from her victim impact statement. Miller's initial proposed quote was rejected by Stanford on the premise that it might be triggering to survivors, \cite{drell_2018} prompting outrage and activism on campus. \cite{knowles_2018} Stanford rejected additional quotes that Miller proposed, and Miller withdrew from the negotiation process. The otherwise completed contemplative garden, located in a remote part of Stanford campus between two undergraduate residences, was therefore left unmarked for several years. Our team learned about these events in fall 2018, and we realized that AR provided a way to bring Chanel Miller's words to the garden and create engagement with the physical space in a way that proved difficult otherwise.

\subsubsection{Stakeholder interviews}

Starting this project in a thoughtful way required identifying and communicating with many stakeholders. We had conversations with students who had lived near the garden, activists who had rallied for a plaque, administrators at Stanford and its Title IX office, and student leaders past and present. Interviewees recommended follow-up interviews with additional subjects. 

We performed semi-structured interviews with past and present Stanford students, discussing their knowledge of and relationship to the garden, their opinions about the plaque, and what they wished to impart on the garden's future visitors. We also collected stories from students who lived near the space, one of whom described overhearing a pair of male students talking about their sexual exploits in the unmarked garden, illustrating the need for a marker at the site. These anecdotes from interviews were edited into clips which formed the audio content of the AR experience.

\subsubsection{Design}

We received design feedback from students, activists, and university stakeholders on an ongoing basis. The final version of the experience began with a brief textual introduction to Chanel Miller, the context behind the garden, and our intentions for the project. The user then began the AR portion of the experience by pointing the tablet viewfinder at an image marker placed next to the garden. When the tablet successfully anchored, two plaques and seven letters materialized on-screen in and around the garden. The plaques were digital renditions of two different physical plaque designs displaying two different quote suggestions from Chanel Miller. Tapping on the plaques played audio recordings of Miller reading the quotes. The plaque at the entrance of the garden, where the original plaque was proposed for installation, featured the first quote that was rejected by Stanford. The second plaque, placed in front of the fountain on the ground, featured another quote Miller suggested but that was also not approved.

AR letter envelopes, with attached audio from student interviewees discussing the space, floated near eye level in the garden during the experience. Tapping a letter opened an audio player which played the corresponding clip. The letters played interviews of various students and student activists, all of whom were asked what they would want future visitors to the garden to know (hence the title, \textit{Dear Visitor}). A counter in the corner of the screen kept track of how many clips the user had heard. See accompanying supplemental materials for a video of an abbreviated walk-through of the experience.

\subsubsection{Tech \& Implementation}

The experience was created in Unity with an Apple ARKit SDK and 3D models created in Blender, and was deployed to iPad Pro tablets. We decided to use tablets instead of headsets to maximize visual fidelity and viewport size while minimizing cost. At the time of the project's deployment, headsets like the Magic Leap were both prohibitively expensive and used optical see-through designs, which meant they had small viewports and struggled in outdoor environments. We found smartphone viewports to be too small for a high-quality experience.

The constraints of the project guided our choice of AR anchoring mechanism. We used image markers, which match an image file in the app to the image's printed counterpart. Marker-based anchoring is common and fairly easy to use, but most importantly, we found this method to be the most reliable in the face of lighting changes and other outdoor environmental factors. We used a printed image, 0.2m by 0.2m, placed next to the garden, which was scanned by each experience participant.

While we extensively considered other anchoring mechanisms such as 3D object recognition (e.g., of inset stones around the fountain), we found them unreliable outdoors. In particular, they had difficulty adapting to changes in lighting even within a single day. Anchoring to a 3D stone pattern during the day might work at first, but if the stone pattern were obscured by shadows from nearby trees a few minutes later, anchoring would fail. If 3D object or scene recognition were to one day become more reliable, it would be the preferred method of anchoring, since it would be able to anchor on features native to the physical scene.

\subsubsection{Launch}

The Dear Visitor experience was painstakingly created and reviewed with stakeholders, a process which took almost an entire academic year. As the experience neared completion, we conducted a beta test, followed by a full launch, which fortuitously coincided with Chanel Miller's book launch and identity reveal (she had previously been known as Emily Doe).

\begin{figure}[h]
  \centering
  \includegraphics[width=3.5in]{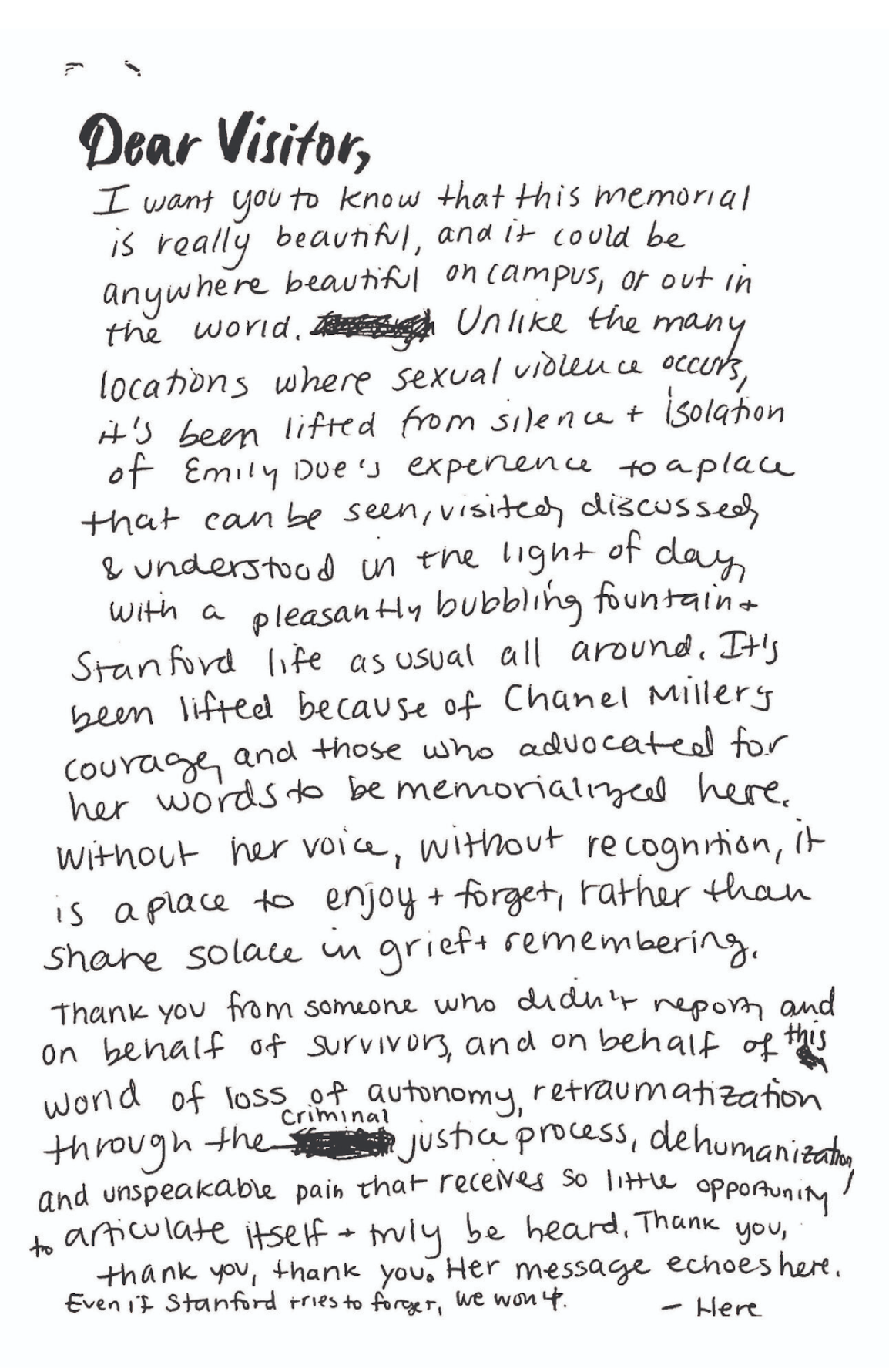}
  \caption{An example anonymous letter written in the offboarding experience of \textit{Dear Visitor}}
  \Description{Scanned image of a white sheet of paper with curly black handwritten script at the top reading ``Dear Visitor." A hand-written letter is scribbled on the paper detailing a participant's experience.}
  \label{fig:offboard}
\end{figure}

We organized a day-long launch event in September 2019. More than 100 people participated in the experience, which had ``ticketed" time slots to control crowding in the space. We had on-site volunteers start users with the tablet and headphones and remain nearby in case of technical difficulties. Because of the charged nature of the subject matter, we designed a voluntary, non-digital offboarding experience to help participants process the app experience. All visitors were invited to participate in the offboarding experience, and most did. Participants were invited to hand write their own physical ``Dear Visitor" letters to others who visit the space. 
Figure~\ref{fig:offboard} shows an example of a letter.
More than 50 such letters were written on the day of the event. To conclude the event, the remaining visitors gathered in a circle to read powerful quotes from Chanel Miller's victim impact statement in solidarity.

\subsubsection{Impact}

Many publications wrote about the \textit{Dear Visitor} launch event given the high-profile nature of Chanel Miller's identity reveal and book release, including \textit{BBC News}, \textit{The Guardian}, and \textit{The LA Times}. This press allowed us, and the many other activists dedicated to this issue, to direct further attention to the issue of the plaque.
Following the launch event and Miller's book release, calls for action to install the real plaque at the garden proliferated. A petition to install the plaque gained more than 2,200 signatures and was sent to the Stanford administration, and the Stanford Faculty Senate unanimously voted to install it. \cite{le_2018} Several unofficial physical plaques were put up at the site by activists during this period. \cite{shao_2019} In February of 2020, the official plaque was installed. \cite{sd_2020}

\begin{figure}[h]
  \centering
  \includegraphics[width=\linewidth]{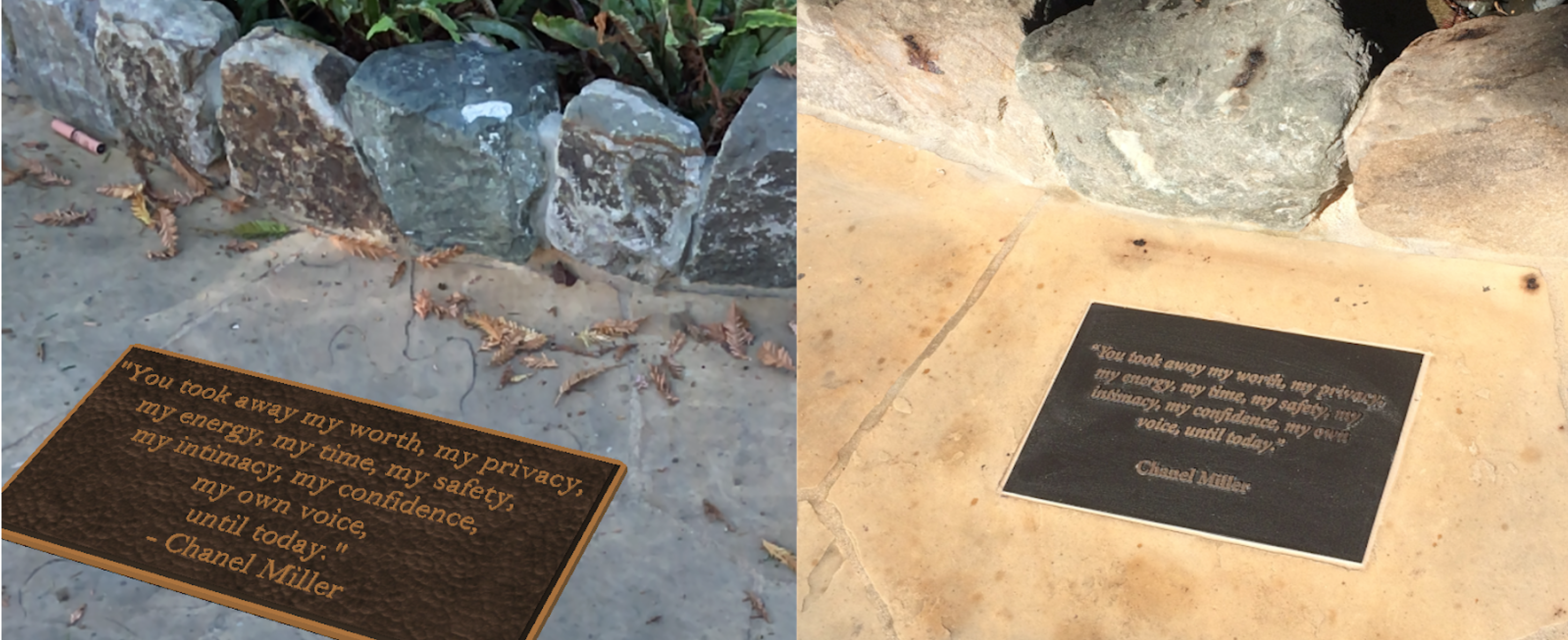}
  \caption{Left: digital plaque from \textit{Dear Visitor} in 2019 overlaid at the garden. Right: installed plaque in 2020, whose design and placement took inspiration from \textit{Dear Visitor}.}
  \Description{On the left is a screenshot of the digital plaque in \textit{Dear Visitor} in 2019 overlaid on the monument site in AR. The plaque reads "You took away my worth, my privacy, my energy, my time, my safety, my intimacy, my confidence, my own voice, until today." On the right is a photograph of the installed physical plaque in 2020, whose design and placement is nearly identical to the overlaid AR plaque on the left.}
  \label{fig:dv_sbs}
\end{figure}

Figure~\ref{fig:dv_sbs} shows a side-by-side image of the digital plaque on the left, designed by us for the \textit{Dear Visitor} experience, and the plaque that was eventually installed onsite on the right. The similarity is noteworthy because neither the digital model nor its placement were part of the original design for the plaque, which featured a longer quote and had a taller form that was slated for the entrance to the garden. We thus have reason to believe the final design of the plaque installed in 2020 took direct inspiration from our digital design, an unexpected outcome of the project. 

\subsection{Charleston Reconstructed}

\subsubsection{Context}

In June 2015, a young white supremacist named Dylann Roof entered Charleston’s oldest historically African American church (Mother Emanuel), joined the congregation in prayer, and then murdered nine of its members in an effort to start a race war. South Carolina’s state legislature agreed that it was time to remove the Confederate flag from its state house, but less than a mile from Mother Emanuel, a 115-foot statue of former Vice President John C. Calhoun, a prominent defender of slavery, still stood in Marion Square. It was emblematic of a problem common to many public spaces in the United States: an outdated monument codifying and celebrating an outdated understanding of the country's history.

Historians Ethan J. Kytle and Blain Roberts claim that ``No place in America has spent as much time and energy selling memories\textemdash most whitewashed, others unvarnished \textemdash of its past" as Charleston, the port city through which 40\% of enslaved Africans passed in the Transatlantic slave trade.
\cite{kytle2018denmark, 40} In the charged physical and historical landscape of the city's central square, we saw an opportunity to contextualize and re-imagine monuments both present and absent. From a technological perspective, we saw an opportunity to bring a physical space into dialogue with its history through AR.

\subsubsection{Final design}

\begin{figure}[h]
  \centering
  \includegraphics[width=\linewidth]{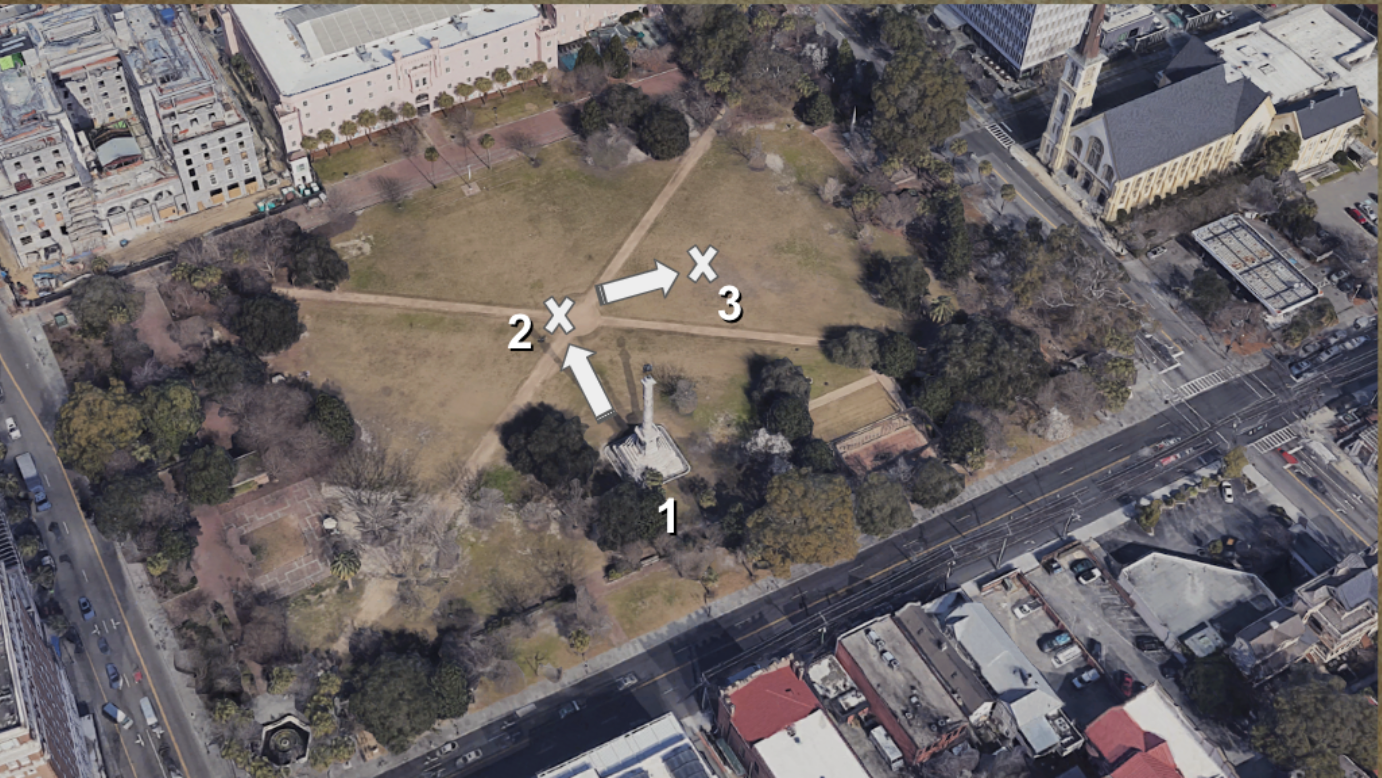}
  \caption{User journey from chapters 1 to 3 in Marion Square}
  \Description{Satellite image of Marion Square from above, which forms a large white "X" across a grassy field with three labeled points (1 to 3) indicating the sites of the three chapters of the experience.}
  \label{fig:square}
\end{figure}

The design of the app took the form of three themed "chapters," meant to be experienced in situ in Marion Square. Users were prompted to explore several sites in the square using a tablet. Tapping on the digital objects that had been digitally overlaid on the physical space triggered interview audio from historians and key community stakeholders. From a content perspective, our goal was first for users to consider the white-supremacist narrative represented by the existing state of the park's monuments; then to consider a counter-narrative centering the contributions of Black Charlestonians; and finally, to consider more broadly how and whether design choices in public spaces can be made to represent a community's shared values and history-- perhaps in a different form than traditional monuments that take the form of a person. Each of the three chapters used a different affordance of AR, allowing for three different ways to re-imagine public spaces more equitably. A map of the user journey from chapters 1 to 3 through Marion Square can be seen in Figure~\ref{fig:square}.

\begin{figure}[h]
  \centering
  \includegraphics[width=\linewidth]{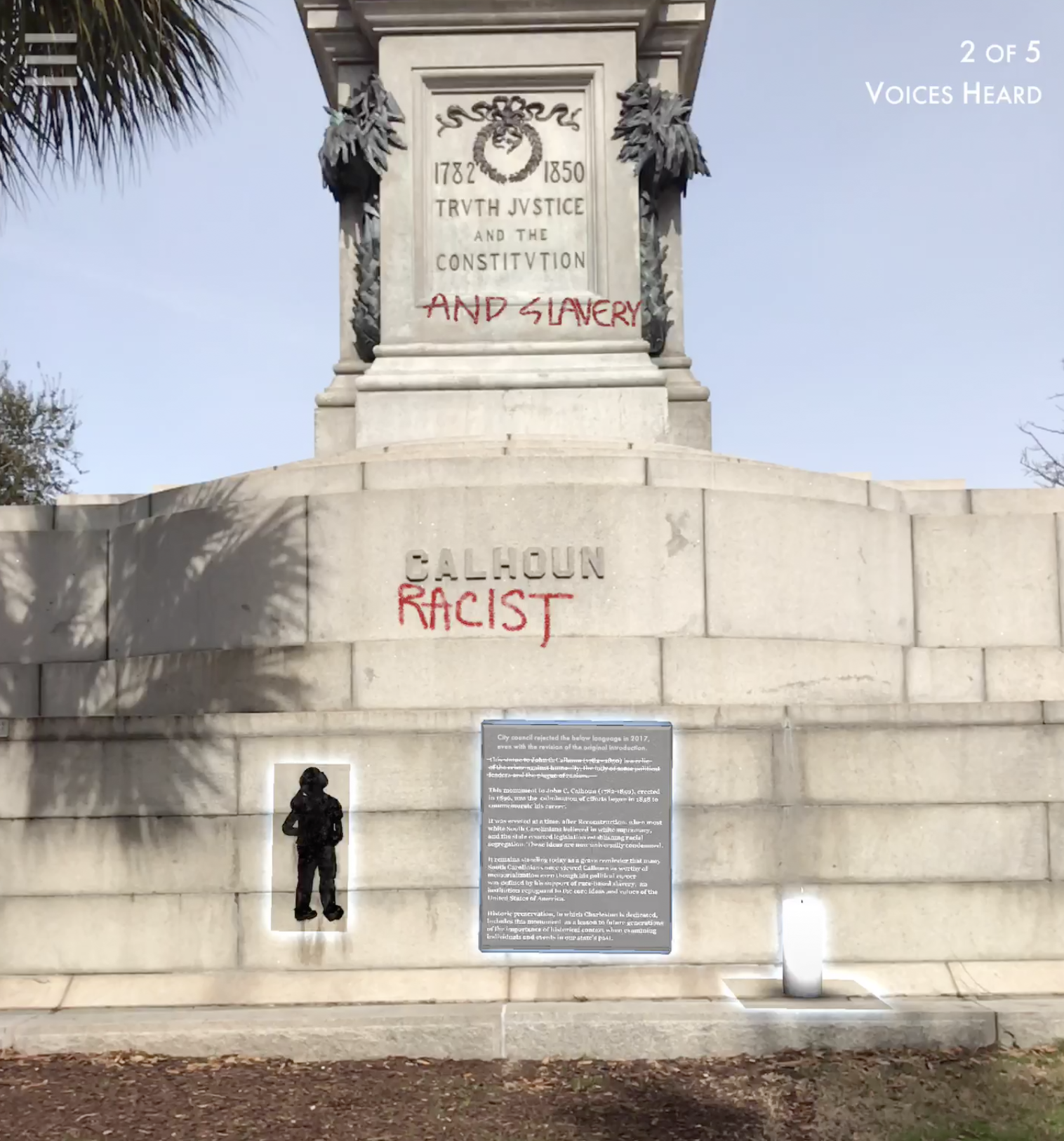}
  \caption{Screenshot from Chapter 1 of \textit{Charleston Reconstructed}. Digitally rendered graffiti, a proposed plaque, a silhouette of a young girl, and a candle glow are shown atop the monument. When tapped, each element triggers audio context on the space.}
  \Description{The base of the stone statue of Calhoun. Under Calhoun's name in red spray paint reads the word ``racist". Under the words "truth, justice, and the constitution" read the words "and slavery" written in red spray paint. At the base of the statue is a digitally overlaid plaque. On the left of it is the digital silhouette of a child looking up at the statue. On the right is a digitally rendered candle object.}
  \label{fig:calhoun}
\end{figure}

In the first chapter, we used AR to digitally alter and contextualize Marion Square's prominent monument to Calhoun. The app directed users to stand in front of the Calhoun statue. Users then saw a digital plaque appear in the space. This plaque, shown in Figure~\ref{fig:calhoun}, displayed language the city considered adding to the statue's base but ultimately tabled due to internal disagreements over the language used to describe Calhoun's role in slavery. Tapping on the plaque triggered the voice of Charleston Mayor John Tecklenburg, who expressed his regret at not being able to have the plaque added to the statue. Simultaneously, we overlaid a reproduction of graffiti that was sprayed onto the monument after the 2015 shooting at Mother Emanuel AME Church. Tapping on the recreation of the spray-painted word, "racist," at the base of the statue triggered the voice of Daron Lee Calhoun II, a Black Charlestonian who noted that Calhoun had once "owned" his family. This chapter was designed to use AR to instantiate small interventions to the existing monument and interrogate its lack of context\textemdash through plaques or other methods\textemdash in the physical world. 

For the second chapter, we instantiated in the square a statue of Denmark Vesey, a native Charlestonian who was executed for plotting what would have been the largest slave insurrection in American history in the early 1800s. This was a digital reconstruction of an existing monument, located elsewhere in Charleston. A committee of historians and activists originally proposed that the Vesey statue be installed in Marion Square in 1996, but many Charlestonians objected, and the Vesey statue was eventually constructed in a park two miles away\textemdash well outside the area commonly visited by tourists. This motivated another novel intervention with AR: using photogrammetry to generate a 3D model of the Vesey statue, then instantiate it at the center of Marion Square. This would allow users to visualize what the square would look like if the statue had been there rather than two miles away.

When users tapped on the Vesey statue in the app, we resized the digital model of the statue so it appeared to ``grow" as tall as the real-life Calhoun statue, as seen in Figure~\ref{fig:vesey}.
Simultaneously, we included audio of community members discussing the need for better representation in Charleston’s statuary and the many African American figures from Charleston’s past who deserve their own monuments.

For the third chapter, we designed and instantiated a digital monument from scratch and overlaid it on the space. Conceptually, this was intended to provide an alternative model of what statues in public space could commemorate. Statues are usually built in dedication to a particular person who embodies community values or aspirations. As shown in recent years, historical figures make for complicated subjects for monuments given the complex nature of any one person's legacy. We took this opportunity to imagine what a monument to a shared set of stories, hopes, and values could look like. From an HCI standpoint, we also designed and instantiated this virtual monument that had features that are impossible in the physical world, like interactive audio combined with physics-defying visual features. 

We commissioned College of Charleston architectural student T'Leya Walker to design a virtual monument representing sentiments gathered in our interviews. Walker's final design incorporated symbolic elements including wrought ironwork in honor of local African American blacksmiths, palmetto trees symbolizing strength and flexibility in the face of hurricane winds and the resilience of Black Charlestonians, and fingerprints in reference to the hands that built Charleston. The monument took the form of a brick archway, representing the Door of No Return, through which many West Africans left their home continent on a forced voyage to the United States. Users could tap chains strung across the archway's entry to hear about issues of ongoing inequality in Charleston. After the audio associated with each chain played, it was animated to break and disappear. After breaking each chain, users were instructed to walk towards the archway, at which point the tablet screen faded to black. Users were then left with the onscreen prompt to look back at the park as it stands and consider how it could be different. 

\subsubsection{Stakeholders}

We took a community-oriented journalistic approach, conducting interviews with historians, locals and activists, and using their expertise and lived experiences to determine the design and narrative elements that became part of the experience. After contacting key experts who had written about the topic of monuments and historical memory locally, we asked each interviewee with whom else it was essential for us to speak, and we ultimately recorded more than 25 interviews. As non-Black reporters and designers, we felt it was important to build trust and welcome as much feedback as possible, so we emphasized transparency about our goals and a willingness to maintain ongoing communication with stakeholders.

The community members included as voices in the \textit{Charleston Reconstructed} experience included professors from the College of Charleston and local military college The Citadel; the poet laureates of both Charleston and South Carolina; a representative of the Charleston chapter of the NAACP; multiple local tour guides and public historians who focused on Black history; a distant relative of John C. Calhoun; several community members who were affected by the shooting at Mother Emanuel; and multiple local activists and advocates working toward the removal of the Calhoun statue from Marion Square.

We also maintained contact with representatives from organizations including the \href{https://independentmediainstitute.org/make-it-right/}{Make It Right Project} and the Charleston Activist Network, and presented iterations of our work at several local advocacy events during the design process. We used the interviews and information gathered at these events to inform an informal participatory design process, whereby insight and feedback from our interviewees directly informed the interim and final designs of the app.

A direct example of this feedback system played out in the first interview conducted for the project. We initially considered digitally "replacing" the Calhoun statue with a statue of a Black Charlestonian, and intended to ask interviewees who they would like to see replace Calhoun. But Christine King-Mitchell, a historic interpreter at Charleston's Old Slave Mart Museum, changed our thinking\textemdash she told us she did not think public monuments should be about putting "one man or woman on a pedestal." Rather, she noted that when she walks around Charleston, she thinks about all the enslaved people who made the bricks that built the city. For her, the fingerprints found in those bricks today serve as a monument to "a people that were always in the shadows, but look at the works they left behind."

\begin{figure}[h]
  \centering
  \includegraphics[width=2.8in]{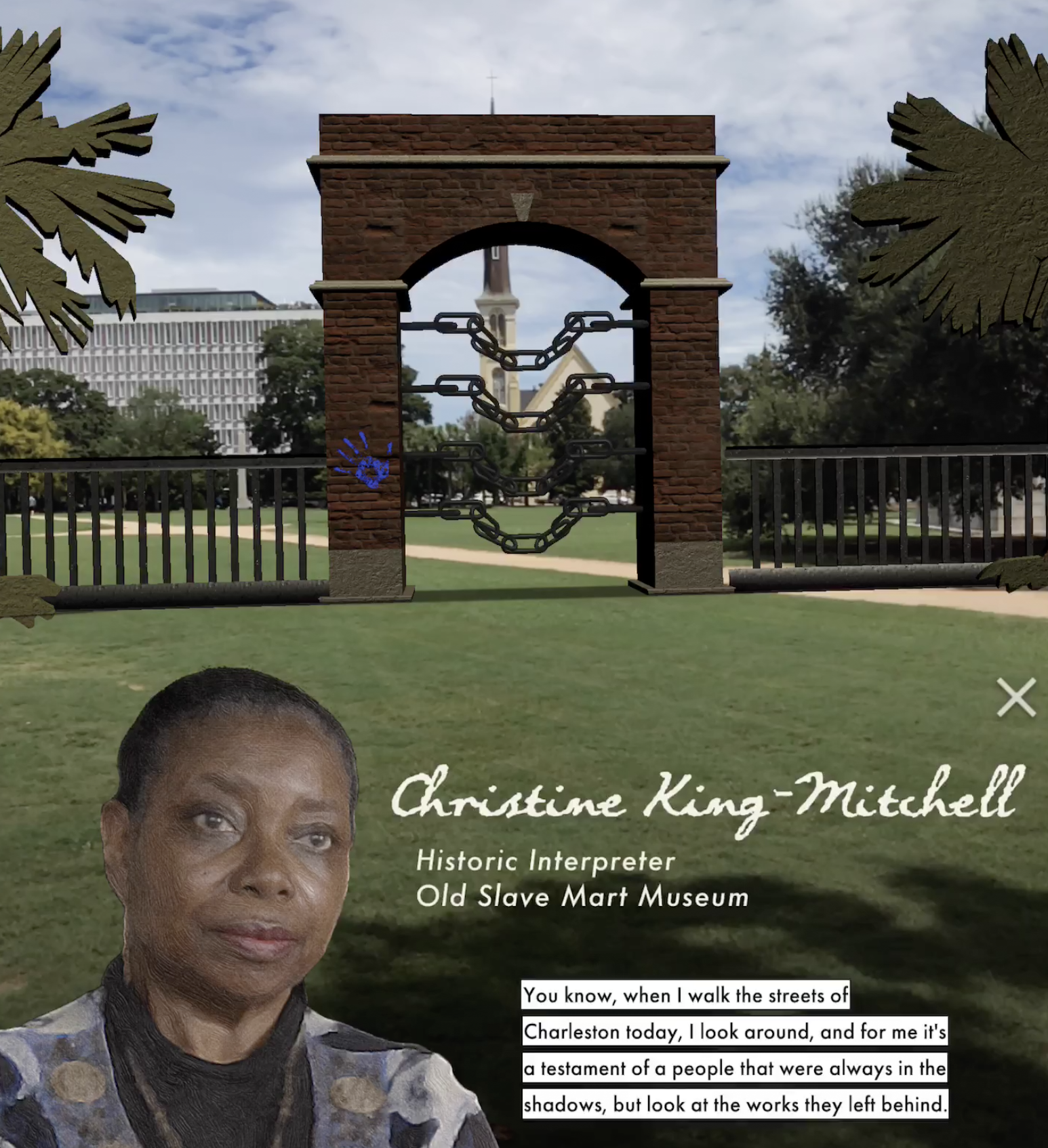}
  \caption{Screenshot of Christine King-Mitchell's narrative in the third chapter of \textit{Charleston Reconstructed.}}
  \Description{A digital brick archway with three chains strung across is overlaid on Marion Square. In the bottom left is a photo of a Black woman, Christine King-Mitchell, a historic interpreter at the Old Slave Mart Museum, and her titles appear under her name to the right of her face. A closed caption from her interview appears under her title, and reads "You know, when I walk the streets of Charleston today, I look around, and for me it's a testament of a people that were always in the shadows, but look at the works they left behind."}
  \label{fig:christine}
\end{figure}

This insight prompted us to consider how we could expand our vision for the application to allow for something broader, more inclusive, and more narratively complex than simply replacing one historic figure with another, and it directly informed the decision to build the the third chapter of the experience in a different way. Her insight led to us to conceptualize the monument in the third chapter as a non-human form made of digital bricks, where tapping a hand-print on the bricks let users hear the perspective King-Mitchell shared with us in that first interview. A screenshot of this final design is seen in Figure~\ref{fig:christine}.

This interplay between community insight and design decisions carried on throughout the project.
Given the sensitivity of the issues at hand and the fact that the AR component was difficult for many participants to understand, we also invited stakeholders to serve as beta testers and to provide feedback prior to any public release of the app. This is a highly unusual practice from a journalistic perspective, but we felt it was essential to collaborate in designing the experience with the communities to whom it was most relevant, especially given the sensitive nature of the topic and our own limited understanding of the Black experience. 

As we identified stakeholder voices for the design process, we were also aware that many people in Charleston wished to preserve the status quo in the park. After interviewing some people with this view, we decided that since the existing treatment of the monument already aligned with their values, there was no need to use AR to represent their vision for the space. For this reason, we did not prioritize the Calhoun monument's defenders as key stakeholders in our design process. In both \textit{Dear Visitor} and \textit{Charleston Reconstructed}, we found that consulting as many stakeholders as possible\textemdash and prioritizing those whose voices may be hard to find or actively pushed aside\textemdash was critical to creating the experience we intended. 

\subsubsection{Implementation}

The setup for \textit{Charleston Reconstructed} was similar to that of \textit{Dear Visitor} (built in Unity with ARKit and deployed to iPad tablets), but the scale of the experience was much larger and complexity of included content was much higher. Designing an AR experience at the scale of a whole public park made it difficult for people to go through the experience on their own, so we implemented guided facilitation to make sure users did not miss visual cues. Some visual cues were built into the experience itself, but tablet displays, while large compared to smartphones, still were too small to consistently facilitate wayfinding for a user without help.

\begin{figure}[h]
  \centering
  \includegraphics[width=2.2in]{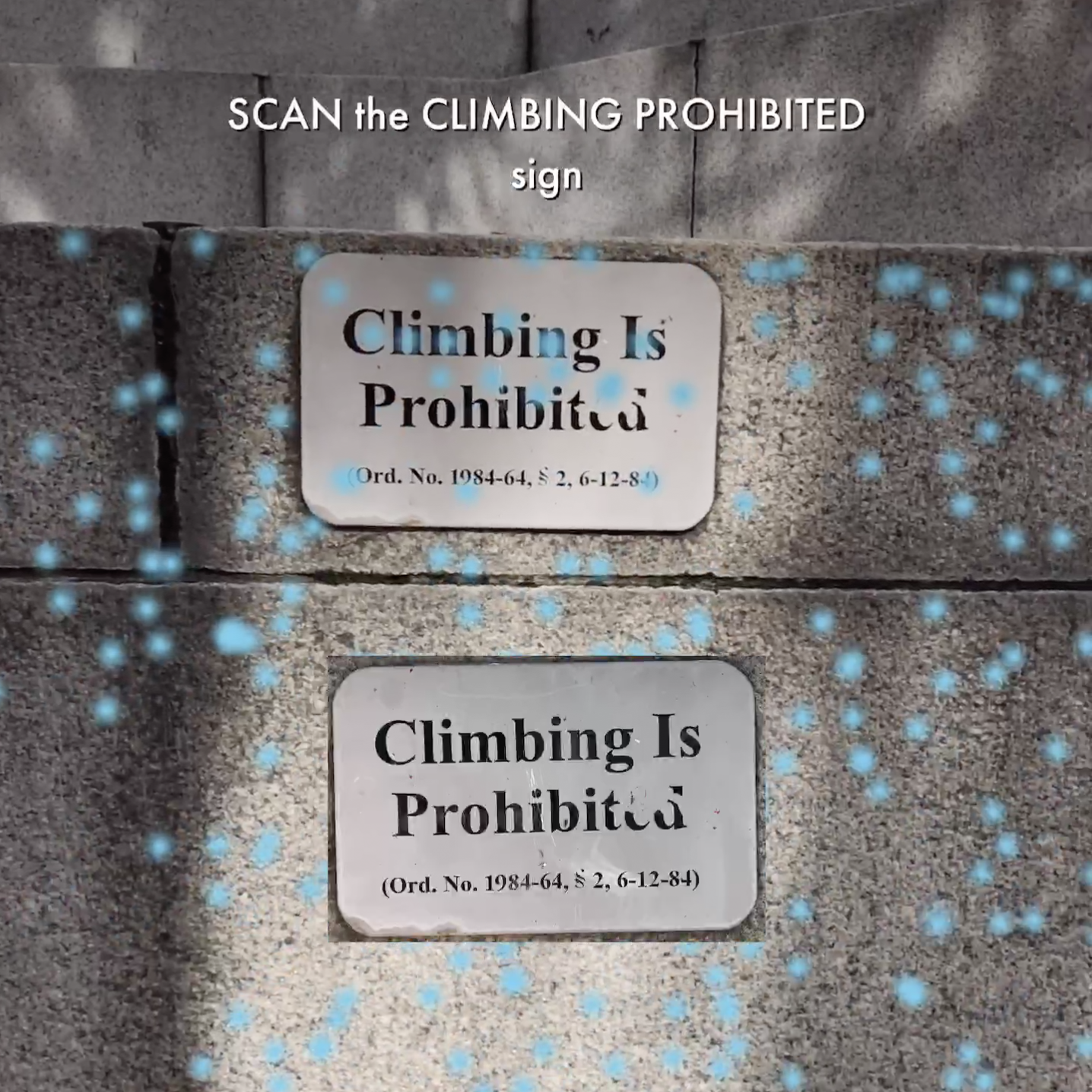}
  \caption{Screenshot of the app, where the prompt was to align an image (bottom) of the ``Climbing is Prohibited" sign at the base of the existing statue with the sign itself (top). }
  \Description{Screenshot of the \textit{Charleston Reconstructed} app, where the prompt was to align an image of the ``Climbing is Prohibited" sign at the base of the existing statue with the sign itself. Blue dots show how the app is searching for the alignment between the  ``Climbing is Prohibited" image it has with the sign in real life.}
  \label{fig:climbing}
\end{figure}

In the first chapter, \textit{John C. Calhoun}, we were able to use marker-based anchoring on an existing marker on the monument: a sign reading "Climbing Is Prohibited," (see Figure~\ref{fig:climbing}) which alleviated the need for additional markers like those needed for \textit{Dear Visitor}.

\begin{figure}[h]
  \centering
  \includegraphics[width=3.0in]{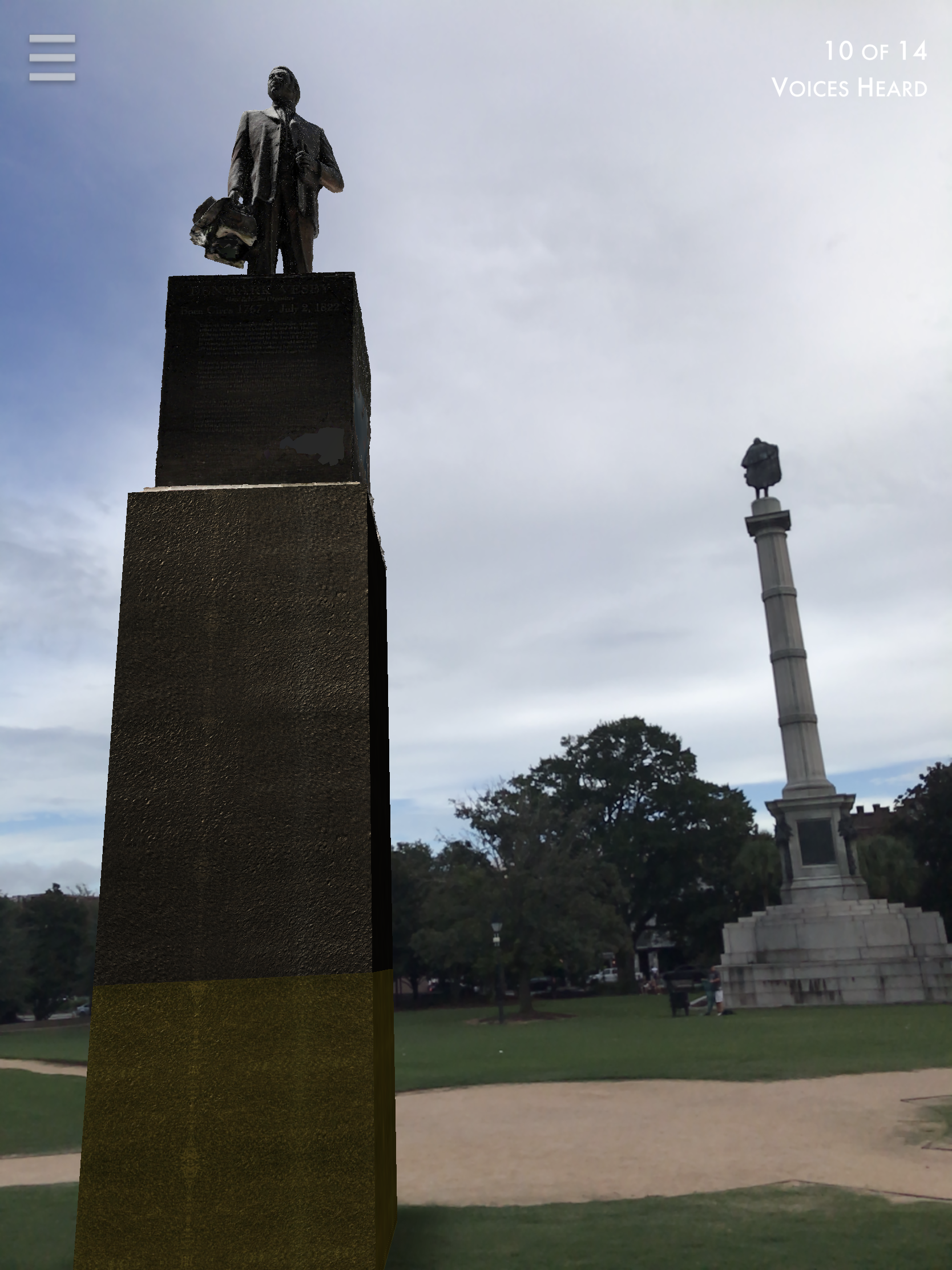}
  \caption{Digitally reconstructed Denmark Vesey statue placed in Marion Square, resized to match the size of the Calhoun statue, seen in the background.}
  \Description{A screenshot of digitally reconstructed statue of Denmark Vesey overlaid in Marion Square, resized to appear as large as the Calhoun statue, which is visible in the background of the image.}
  \label{fig:vesey}
\end{figure}

For the second chapter, we scanned the existing Denmark Vesey statue, located 2 miles away from Marion Square, using drones. We reconstructed the statue digitally using the photogrammetry software RealityCapture, which rendered a 3D model. The statue can be seen as it appeared in the app in Figure~\ref{fig:vesey}.

For the third chapter, T'Leya Walker created a 3D model of a virtual monument in SketchUp. We added interactive elements to this model in order to trigger audio elements in the app and give the chains the appearance of movement.

\subsubsection{Beta test}

In September 2019, we organized a day-long beta test of the experience with about 20 people, most of whom had been interviewed for the project. We scheduled walk-throughs in groups of 2-3 people every hour so that at least one of our team members could personally facilitate for each experience participant. This allowed us to help with any technical difficulties participants encountered, observe how they navigated the application, and note any editorial concerns that arose in real time.

In keeping with the collaborative process we established during the early phases of the project, we solicited ongoing feedback from these participants throughout and after the experience, with the goal of using their notes and our observations to make improvements. This provided crucial insight both technically and editorially about how people respond to and interpret AR in outdoor settings, on which we elaborate in a later section, "\nameref{open_challenges}."

Most notably, we observed that to complete all three chapters and hear all of the included audio clips, which totaled about 43 minutes, each person took approximately an hour and 15 minutes during the beta test. We reduced the number of audio clips and AR elements by nearly half as a result of participant feedback. Following this beta test and edit process, we held several smaller walk-throughs with a few key stakeholders who had emerged as informal advisors due to their interest in the project, all of whom found the changes we made to be effective. We felt at this point that we were ready for a wider public release, and in February 2020 we held our final demo of the experience with a representative for Spoleto Festival USA, who expressed intent to tie our public launch event in with the festival's programming in June 2020.

\subsubsection{Aftermath}

In February 2020, news of a global pandemic halted travel plans. Our planned deployment partner, Spoleto Festival USA, cancelled its in-person festival in Charleston, which was slated for June 2020. In May 2020, George Floyd was killed by police in Minneapolis, Minnesota, launching a nation-wide reckoning with ongoing racism, anti-Black police violence in the US, and the legacy of racism in the US. 

In response to a newly re-energized advocacy and protest campaign that followed in Charleston, Mayor John Tecklenburg took a corrective action that months earlier he said he did not have the power to do: he ordered the John C. Calhoun statue's removal from Marion Square, and it was removed on June 23, 2020.

Overnight, the landmark our experience was built around had disappeared from the park, and with it, the most tangible impetus for the \textit{Charleston Reconstructed} project. It was a celebratory outcome and a previously unthinkable step in the right direction for a city that has historically resisted such changes. And for our team, it meant that doing a full, longer-term, and public launch of the project would have required a near-total rethinking of what we had already built. As such, the project was not launched in full.

\section{Discussion}

\subsection{Practical learnings}

Designing, developing, and deploying these two projects yielded deep insight into the unique logistical challenges of location-based AR, particularly when dealing with social justice issues. We present practical learnings based on technological and spatial constraints to provide guidance for future designs.

\subsubsection{Connecting content and place}

A key editorial decision we made early on was to have the content maintain focus on the physical space as much as possible. This was in an effort to spatially connect the content with the context, justifying the use of AR at a particular site as opposed to a non-spatial medium.

In \textit{Dear Visitor}, we specifically asked each student interviewee to impart thoughts for future visitors to the garden. Some of the most impactful moments of the \textit{Dear Visitor} experience, according to feedback we collected, were in response to this framing, such as when one interviewee made a reference to a building "to your right," and the user could look to the right and see the building. Another audio letter shared the story of a student looking down at the site from her dormitory, a building which was in view for users participating in the \textit{Dear Visitor} experience.

The technical and logistical complexities involved in working with AR as a medium are only worthwhile if the experience's content could not be experienced or understood another way. As such, we cut content that did not need to be experienced in the space, and instead hosted it on our project websites for additional contextual reading. This allowed us to make editorial choices about the experiences without totally sacrificing the voices and stories we believed were important but less critical to hear in the space itself. We could have had a heavier hand in doing this for the Charleston project, which was still quite long.

We found a location-centered framing to be especially effective in showcasing the value of AR, by not only making the content more relevant, but by changing the relationship a user has with the space around them. Even for a user without existing personal connections to the content being shared, the content is made relevant to them by virtue of the fact they are physically standing in the space being referenced.

\subsubsection{Length}

The ideal length of an AR experience is much shorter than we had anticipated. This was especially an issue for \textit{Charleston Reconstructed}, where the initial version contained 43 minutes of content, and took each person approximately an hour and 15 minutes to complete during the beta test. Holding up a device even in the heat and glaring sun (which reflected off the tablet screen) takes physical effort, even for a few minutes at a time outdoors. Many users found it difficult to stay engaged for the duration of the experience, despite finding the chosen interviews and the AR components powerful and engaging.

We ultimately cut the length and number of audio clips down significantly based on feedback, with \textit{Dear Visitor} and \textit{Charleston Reconstructed} ending up at 13 and 16.5 minutes of content respectively. This was still likely longer than ideal, but ultimately we had to balance the total length with a thorough and nuanced treatment of the subject matter. We also added a counter showing how many voices users had heard so far and how many were left in each chapter so that users could make informed decisions about whether they wanted to move on or keep listening. We recommend that future creators strive to keep AR experiences short, ideally not exceeding 5 minutes, or to break up experiences in ways that can be understood in shorter chunks over time. This can be difficult for social justice-related projects that are inherently complex since it can feel reductive to edit down the content. This challenge must be weighed against the risk of losing AR users entirely given natural focus limits, especially in challenging outdoor environments.

\subsubsection{Onboarding and orienting users}

We found that users needed clear, simple instructions at the beginning of the experience about how to use the application and what they should expect. During the beta test for \textit{Charleston Reconstructed}, we often needed to provide verbal instructions or borrow the tablet from users briefly to demonstrate what to do in order for them to begin the first chapter successfully. We fixed this after the beta test by building in a prologue that prompted users to perform all of the basic actions that would be required of them throughout the remainder of the experience before getting to the first chapter. Subsequent walk-throughs with smaller groups confirmed that priming users more directly at the beginning allowed them to navigate the app more quickly and successfully.

For \textit{Charleston Reconstructed}, users needed clear instructions to orient themselves within the park before launching each chapter given the large size of the site. The beta version directed users via onscreen text and included a button to press when they had walked to the correct location. However, many users failed to interpret the button, which read “I am here,” as intended. We fixed this by adding a map of the park displaying where to go, a photo of what the viewer should be seeing when they were in the right place, and we changed the button’s text to say “Tap here when your view matches the photo below.”

\subsubsection{Human facilitation}

Despite implementing better onboarding for \textit{Charleston Reconstructed} after our first beta tests, we still found that human facilitation was critical to the success of the user experience, especially for less tech-savvy users. Even in a small garden like the one at Stanford in \textit{Dear Visitor}, a human facilitator often helped users get back on course in the event of a technical difficulty or a question from a participant. This was true both in our beta tests as well as our final launch. Having ready facilitators meant that technical challenges or discomfort were less of a barrier to the participant meaningfully engaging with the content of the experience. In a space as large as Marion Square in Charleston, and with the generally less tech-savvy users we worked with there, human facilitation in order to create a smooth experience of the content was even more important. The level of human facilitation we found to be crucial to the successful delivery of these experiences came at the expense of their scalability. This finding is revisited as an open challenge for this kind of work.

\subsubsection{Closed captioning}

In initial iterations of \textit{Charleston Reconstructed}, closed captioning was persistently onscreen, accompanying each new piece of audio that was triggered. We found closed captions discouraged participants from keeping their eyes on the main AR components after tapping on them to trigger audio elements, because users would instinctively read along with the captions. We fixed this by making closed captioning optional in the final version for those who needed it. For \textit{Charleston Reconstructed}, we observed that the sounds of a heavily-trafficked park could be a distraction to some users, and found that adding music and ambient sounds actually helped users immerse themselves in the app better than operating it in silence. The general applicability of this finding may depend on the nature of the site of future work.

\subsubsection{Content placement}

The beta version of \textit{Charleston Reconstructed} included a series of symbols running up the column of the John C. Calhoun statue, but it was difficult to keep them anchored properly so high above the ground. It was also not intuitive for users to point the tablet up high at the sky to see these elements. It was not ergonomic to hold the tablet up high for so long, and sunlight was often a source of irritation. We eventually removed most of these skyward visual elements and kept the AR content closer to eye-level so it was easier for users to see and interact with. As such, we learned that the content's placement deeply affected users' interest in interacting, a finding we hope can inform future AR design.

\subsection{Open challenges}
\label{open_challenges}
The fact that these two projects were tied to volatile social issues introduced unique opportunities and conceptual challenges for designing a successful AR experience at each site. 

\subsubsection{Timing}

The timeline for creating and launching these projects ultimately decided each project's fate. For both projects, editorial choices were very difficult, and the projects' content passed through many rounds of validation with key members of each community before being included in the AR experience. For \textit{Charleston Reconstructed} in particular, we did not sacrifice careful judgment for speed in these decisions. This carefulness, along with the added constraint that we had to travel in order to deploy and receive feedback on it, meant the project took nearly two years to develop. In unforeseeable strokes of fate, by the time the project was ready to launch, the pandemic had come, and the main statue around which it centered came down soon after. Because major world events outpaced our development process, the project missed its moment.

On the other hand, the site of \textit{Dear Visitor} was in the town where some of us lived, and some of us were still in the community of the university. This made prototyping onsite and meeting spontaneously with stakeholders to receive feedback much easier. Feedback from the varied stakeholders in the project, including Chanel Miller herself (through a representative), still took a great deal of time, even with the project's much more concise scope than \textit{Charleston Reconstructed}'s. \textit{Dear Visitor} took just under a year from conceptualization to final launch. Being able to launch the same week as Chanel Miller's book allowed us to synergize with existing momentum on the issue of the plaque, and the real plaque was installed a few months after our project launched.   

In both projects, technical and social challenges related to the work contributed to a long time horizon. It took time to understand the problem, consult with experts, and design content to further the goal of creating a truly meaningful AR experience, rather than something that merely engaged with the issue at a shallow level. We urge future creators of AR projects with storytelling on social issues to plan for an extended timeline. If the social issue is ``current" or volatile, the timeline of the project may change in unexpected ways and affect the project's ability to launch. It is worth considering how the project can live on despite potential changes to the space that might be inherent to working on volatile social issues.

\subsubsection{Persistence}

Several major challenges for this kind of work deal with persistence. Creating location-based AR activations on contested public space means that designers should expect that the physical space could change suddenly and dramatically. As we saw in the case of the Calhoun monument being taken down almost overnight after many considered it a lost cause, these changes can be extreme, and can instantly change the context for any AR interventions on top of them. The issue of AR persistence as a project site itself changes will continue to be a challenge for location-specific AR.

As previously discussed, we used visual markers to make up for AR systems' lack of semantic understanding of space. If AR one day achieves more generalized location-based semantic understanding, it will need to be resilient to or account for change in a location's appearance in order to persistently host this kind of experience.

The question of persistence of the technology itself presents its own set of challenges, and the technology's lack of persistence can affect the longevity and impact of the work. AR is changing quickly, and the best technology today will likely be obsolete sooner than we wish. Maintaining our projects as technology changed was difficult even in the short time span during which we worked. Projects working to facilitate long-term collective memory should actively future-proof as much as possible, but as of now, there is limited ability to effectively do so given lack of general standards across which to port AR experiences across platforms.

The challenge of rapidly changing technology also relates to the platforms that host AR experiences. Though any one AR experience-hosting platform has not yet emerged as dominant as of this writing, the private, commercial nature of app  platforms to this point introduces future challenges of ownership and profit over the stories hosted in AR. Given Charleston's massive tourism industry and the prevalence of walking tours, we made plans to pass the finished experience off to local Black tour guides, who could facilitate for users post-launch. We made the decision that any profit gained in the future from \textit{Charleston Reconstructed} would go back to the Black community in Charleston. Protecting stories like the ones shared in these experiences from exploitation is made challenging by the inextricable involvement of corporations in AR development to this point.

\subsubsection{Accessibility}

The persistence of AR technology itself can also affect another area we consider an open challenge: the accessibility and scalability of this kind of experience. Hosting in-person launch events showed us how powerful going through the experience with others could be. Processing difficult historical events in community is a different experience from using an app alone. We opted for human facilitation of \textit{Dear Visitor} and \textit{Charleston Reconstructed} due to the technical challenges inherent to the medium that could be better navigated by dedicated onsite help. Secondly, we found there was immense power in going through the experience and offboarding in community, findings supported by work that shows virtual experiences can significantly affect social connectedness. \cite{miller2019social} The decision we made to keep these experiences facilitator-led and on large iPad screens meant keeping them off app stores, as we did not feel the unfacilitated experience, particularly on a small smartphone, could be consistently high-quality. Requiring human facilitation had the unfortunate consequence that the voices we worked hard to curate specifically for the space can rarely be experienced in situ. \textit{Dear Visitor} had a successful launch, but is not currently accessible for viewing at the site except upon request. Most material is available on the project \hyperref[https://www.dearvisitor.app/]{website}, a digital format that is easier to maintain than AR and for which we can guarantee a consistently high-quality experience. Wider accessibility and greater longevity of complex AR experiences could come through facilitation partnerships with museums or local organizations, a model which would require strong coordination and more resources. The charged and complex nature of the content we were committed to presenting in the projects guided our distribution decisions, but accessibility trade-offs may have been different if we had designed simpler experiences with content of a different nature.

Without adding in permanent physical markers to a site that refer to the AR experience, there is also limited opportunity for a person walking through a space to ``discover" the additional knowledge that might be gained by doing the AR experience, decreasing the experience's reach. On the other hand, AR applications can facilitate an interesting design feature: people can opt in to content that all people may not wish to (or be able) to see. The issue of sensitive content in \textit{Dear Visitor} was minimized with an explicit content alert at the beginning of the experience, which addressed concerns for the university and sexual assault experts about how the content could affect survivors, students, and casual viewers. Given ongoing societal conversations about content warnings and how to present sensitive content in a responsible way, site-specific AR applications provide an interesting opportunity to build in consent features before viewing certain kinds of content. AR's current accessibility and discoverability challenges thus give it unique affordances that differ from purely physical changes to challenging sites: while site-specific AR does not currently lend itself to an easily discoverable, accessible shared experience, the technology can allow people to opt in to layers of additional content as they see fit.

\subsubsection{Anchoring to the space}

For \textit{Dear Visitor}, we decided to anchor AR content through a printed image marker since it was the most resilient tracking method. We prioritized reliability first and foremost, knowing that technical difficulties would easily distract from the content. Although we were fortunate enough to be able to use a pre-existing marker on a statue base for \textit{Charleston Reconstructed}, in the future, all location-based AR experiences would ideally be accessible onsite and without additional markers. In the future, AR systems could incorporate location and surface information with world knowledge, aligning an AR experience on the physical landscape despite surface changes in the space without the use of image markers.

\subsubsection{Content curation}

We initially considered creating an open comment platform allowing responses from the public in reaction to our projects, but we eventually decided to heavily curate the content due to the sensitive nature of the topics at hand. Currently, social reactions to digital interventions in public space are not typically captured. Balancing a desire to seek and reflect community input with the realities of difficult, time-intensive content moderation is an open challenge for collaborative digital projects, including those which may use AR as a technology and involve social justice challenges where many community members have opinions and stories.

Related to the challenge of potentially curating comments about the project lies another conceptual challenge: who has the right to tell stories in public space? Narrative control over public space has usually been dominated by those in power, and these projects sought to add digital layers that surfaced stories at risk of being forgotten or suppressed. It remains to be seen how adding digital stories to physical space would play out at scale. Will digital memories in public space compete for attention by volume? Who will curate them over time? What control will the platforms that host them have in their curation and monetization? 

In parallel, the sudden ubiquity of text-to-3D models like Google's DreamFusion, \cite{poole2022dreamfusion} which creates 3D models using a text prompt, makes creating and distributing digital objects easier than ever. As we saw in \textit{Dear Visitor}, digital objects can easily be made manifest in real life, pointing to an additional set of potential implications for the real world as digital designs proliferate. 
These are conceptual open challenges with which designers of AR experiences must engage as the technology becomes more accessible.


\section{Conclusion}

We believe that AR is best used to connect digital content with the context in which it is presented. In our projects, we did this by linking digital layers of AR objects and stories to a contested physical monument in public space. A generalized technical solution for this kind of project is not yet available, but our two location-based prototypes \textit{Dear Visitor} and \textit{Charleston Reconstructed} explored the potential of AR to meaningfully engage with contested public spaces, highlighting underheard stories and proposing visual changes using digital objects along the way.

We found that what we build in AR has real stakes, because virtual content facilitates real experiences, and because it can influence tangible changes to the physical world that expand impact beyond those who witnessed the AR experience.

We intended for our projects to add value and meaning to the world on their own terms, in addition to any changes they might spur in physical space at their respective sites. To go beyond the affordances of the physical world, we added interventions in AR that are not accessible or possible in physical space alone at a monument site.
Examples of the interventions specific to AR in \textit{Charleston Reconstructed} included "moving" a monument from another part of the city, recreating an AR representation of graffiti that was long ago removed from Marion Square, and adding a new, physics-defying digital AR monument. As such, the virtual experiences we designed created layers of meaning on top of what the physical space on its own could present.
David Chalmers argues that virtual objects are real objects, \cite{chalmers2017virtual} and it follows that interacting with virtual objects gives rise to real, compelling experiences\textemdash not just simulations of experiences that then necessarily translate into the physical world. We can attest to the power of AR experiences that add dimension to our non-digital lives based on the reactions we received to the launch of \textit{Dear Visitor} and the soft launch of \textit{Charleston Reconstructed}.

Though we imbued the projects with meaning regardless of changes they might inspire to physical space, by chance, we found strong evidence that what is designed and shown in the digital AR world can, in fact, manifest in the physical world. This occurred when our design and placement of the digital plaque in 
\textit{Dear Visitor} was used for the physical plaque installation at the garden site at Stanford. AR's potential to inspire a jump from a virtual design to a real physical change is exciting. In a context as high-stakes as the conflict over public monuments, it also comes with great responsibility, and with implications that should be taken seriously by those who design these kinds of digital objects and experiences.

Severe political polarization in the US and beyond is often fueled by increasingly siloed digital echo chambers, but public space is something we still have in common and will continue to share. Enduring links between the digital and physical worlds are possible, and AR is uniquely positioned to facilitate new kinds of connections and experiences in shared space that bring people together if done thoughtfully, and if the technology can consistently support them. How the HCI community engages with the connection between the design of AR and public space will have major consequences for society on social justice issues and beyond.

%
\begin{acks}
We extend our thanks for the funding and support from the Brown Institute for Media Innovation. We thank Mark Hansen, Maneesh Agrawala, Ann Grimes, Michael Krisch, Juan Francisco Saldarriaga, and Amy Menes. We also thank Jeremy Bailenson, Geri Migielicz, Shelley Correll, and Alison Dahl Crossley at Stanford. We thank Michele Dauber and Chanel Miller. We thank Matt Shimura, Chanel Kim, Angela He, and Barna Szász for their work, feedback, and support of this project.
We thank the many scholars, artists, community members and activists whose time and insights were shared in Charleston. We thank T'leya Walker for her contributions to this project. We thank Gail Mazzara and Kent Coit. We thank Kali Holloway and the Make It Right Project.
\end{acks}

\bibliographystyle{ACM-Reference-Format}
\bibliography{sample-base}

\appendix


\end{document}